\documentclass[11pt,a4paper]{article}
\usepackage[svgnames]{xcolor}
\usepackage{amsmath,amsfonts,amstext,amssymb,txfonts,pifont,url}
\usepackage[normalem]{ulem}

\usepackage{graphicx}

\usepackage{tikz}
\usetikzlibrary{shapes,arrows}
\tikzstyle{block} = [rectangle, inner sep=1ex, draw]
\tikzstyle{loopblock} = [block, line width=1.5pt]
\tikzstyle{dashedblock} = [block, dashed]
\tikzstyle{edge} = [draw,->]
\tikzstyle{dashededge} = [edge, dashed]
\tikzstyle{boldedge} = [edge, thick]
\tikzstyle{edge1} = [edge, densely dashed]
\tikzstyle{edge2} = [edge, densely dotted]
\tikzstyle{edgelabel} = [fill=DarkOrange, circle, draw=black, thin, inner sep=0mm]

\newcommand{\sampleline}[1]{\protect\tikz\protect\draw[#1] (0pt,0pt) -- (2em, 0pt);}

\usepackage{listings}
\usepackage[latin1]{inputenc}
\usepackage[T1]{fontenc}
\usepackage{url}
\usepackage{hyperref}

\usepackage{algorithm}
\usepackage{algpseudocode}
\usepackage{ifpdf}
\usepackage{verbatim}
\usepackage{subfigure}

\renewcommand{\epsilon}{\varepsilon}
\usepackage{paralist}
\usepackage{xspace}
\usepackage{float}

\newcommand{\true}{\mbox{true}}
\newcommand{\false}{\mbox{false}}

\newcommand{\widening}{\mathop{\triangledown}}
\newcommand{\bbQ}{\mathbb{Q}}
\newcommand{\bbZ}{\mathbb{Z}}
\newcommand{\parts}[1]{\mathcal{P}(#1)}
\newcommand{\abstr}[1]{#1^\sharp}

\newcommand{\aspic}{\textsc{Aspic}\xspace}

\newcommand{\ite}[3]{\textit{ite}(#1,\allowbreak #2, \allowbreak #3)}

\newcommand{\stressalgo}[1]{\colorbox{DarkOrange}{#1}}

\begin{document}
\title{Using Bounded Model Checking \\ to Focus Fixpoint Iterations
\thanks{This research was partially funded by ANR project \href{http://asopt.inrialpes.fr}{``ASOPT''}.}}
\author{David Monniaux \thanks{CNRS, VERIMAG, Gières, France}
\and Laure Gonnord \thanks{Universit\'e Lille 1, LIFL, Villeneuve d'Ascq, France}}

\maketitle

 \begin{abstract}
Two classical sources of imprecision in static analysis by abstract interpretation are widening and merge operations. Merge operations can be done away by distinguishing paths, as in trace partitioning, at the expense of enumerating an exponential number of paths.

In this article, we describe how to avoid such systematic exploration by focusing on a single path at a time, designated by SMT-solving. Our method combines well with acceleration techniques, thus doing away with widenings as well in some cases. We illustrate it over the well-known domain of convex polyhedra.

 \end{abstract}

\section{Introduction}
\label{sec:intro}
\label{part:motivations}

Program analysis aims at automatically checking that programs fit their specifications, explicit or not --- e.g. ``the program does not crash'' is implicit.
Program analysis is impossible unless at least one of the following holds: it is unsound (some violations of the specification are not detected), incomplete (some correct programs are rejected because spurious violations are detected), or the state space is finite (and not too large, so as to be enumerated explicitly or implicitly).
\emph{Abstract interpretation} is sound, but incomplete: it over-approximates the set of behaviours of the analysed program; if the over-approximated set contains incorrect behaviours that do not exist in the concrete program, then false alarms are produced. A central question in abstract interpretation is to reduce the number of false alarms, while keeping memory and time costs reasonable~\cite{ASTREE_PLDI03}.

Our contribution is a method leveraging the improvements in SMT-solving to increase the precision of invariant generation by abstract fixpoint iterations. On practical examples from the literature and industry, it performs better than previous generic technique and is less ``ad-hoc'' than syntactic heuristics found in some pragmatic analyzers.

The first source of imprecision in abstract interpretation is the choice of the set of properties represented inside the analyser (the \emph{abstract domain}). Obviously, if the property to be proved cannot be reflected in the abstract domain (e.g. we wish to prove a numerical relation but our abstract domain only considers Boolean variables), then the analysis cannot prove it.

\begin{lstlisting}[float,caption={C~implementation of $y=\sin(x)/x-1$, with the $-0.01 \leq x \leq 0.01$ range implemented using a Taylor expansion around zero in order to avoid loss of precision and division by zero as $\sin(x) \simeq x \rightarrow 0$.},label=lst:sinc_minus1]
if (x >= 0) { xabs = x; } else { xabs = -x; }
if (xabs >= 0.01) {
  y = sin(x) / x - 1;
} else {
  xsq = x*x;  y = xsq*(-1/6. + xsq/120.);
}
\end{lstlisting}

In order to prove that there cannot be a division by zero in the first branch of the second if-then-else of Listing~\ref{lst:sinc_minus1}, one would need the non-convex property that $x \geq 0.01 \lor x \leq -0.01$. An analysis representing the invariant at that point in a domain of convex properties (intervals, polyhedra, etc.) will fail to prove the absence of division by zero (incompleteness).

Obviously, we could represent such properties using disjunctions of convex polyhedra, but this leads to combinatorial explosion as the number of polyhedra grows: at some point heuristics are needed for merging polyhedra in order to limit their number; it is also unclear how to obtain good widening operators on such domains.
The same expressive power can alternatively be obtained by considering all program paths separately (``merge over all paths'') and analysing them independently of each other. In order to avoid combinatorial explosion, the \emph{trace partitioning} approach \cite{Rival_Mauborgne_TOPLAS07} applies merging heuristics. In contrast, our method relies on the power of modern SMT-solving techniques.

The second source of imprecision is the use of \emph{widening operators}~\cite{CousotCousot_JLC92}. When analysing loops, static analysis by abstract interpretation attempts to obtain an \emph{inductive invariant} by computing an increasing sequence $X_1,X_2,\dots$ of sets of states, which are supersets of the sets of states reachable in at most $1,2,\dots$ iterations.
In order to enforce convergence within finite time, the most common method is to use a widening operator, which extrapolates the first iterates of the sequence to a candidate limit. Optional narrowing iterations may regain some precision lost by widening.

\paragraph{Illustrating Example}
\label{e:g}
\begin{lstlisting}[float,caption={Circular buffer indexing},label=lst:circular]
int x = 0;
while (true) {
  if (nondet()) {
    x = x+1;
    if (x >= 100) x = 0;
} }
\end{lstlisting}

Consider Listing~\ref{lst:circular}, a simplification of a fragment of an actual industrial reactive program: indexing of a circular buffer used only at certain iterations of the main loop of the program, chosen non-deterministically. If the non-deterministic choice \lstinline|nondet()| is replaced by \lstinline|true|, analysis with widening and narrowing finds $[0,99]$.  Unfortunately, the ``narrowing'' trick is brittle, and on Listing~\ref{lst:circular}, widening yields $[0,+\infty)$, and this is not improved by narrowing!
\footnote{On this example, it is possible to compute the $[0,99]$ invariant by so called ``widening up-to'' \cite[Sec.~3.2]{Halbwachs_CAV93},  or with ``thresholds'' \cite{ASTREE_PLDI03}: essentially, the analyser notices syntactically the comparison $x < 100$ and concludes that $99$ is a ``good value'' for $x$, so instead of widening directly to $+\infty$, it first tries~$99$. This method only works if the interesting value is a syntactic constant.}
In contrast, our semantically-based method would compute the $[0,99]$ invariant on this example by first \emph{focusing} on the following path inside the loop:
\begin{lstlisting}[caption={Example focus path},label=lst:focus_path]
assume(nondet());  x = x+1;  assume(x < 100);
\end{lstlisting}
If we wrap this path inside a loop, then the least inductive invariant is~$[0,99]$. We then check that this invariant is inductive for the original loop.

This is the basic idea of our method: it performs fixpoint iterations by focusing temporarily on certain paths in the program. In order to obtain the next path, it performs bounded model checking using SMT-solving.


\section{Background and Notations in Abstract Interpretation}
\label{sec:background}

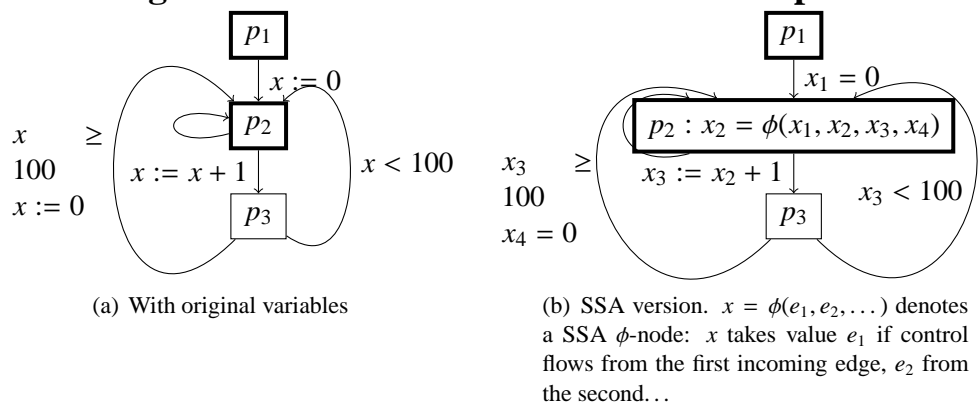
\begin{figure}[htpb!]\vspace*{-10mm}\par%
\subfigure[With original variables]{\label{fig:listing1_cfg}%
\vspace*{-30mm}\par
\begin{tikzpicture}[node distance=1.2cm]
\node [loopblock] (init) {$p_1$} ;
\node [loopblock, below of=init] (loop) {$p_2$} ;
\node [block, below of=loop] (nondet) {$p_3$} ;
\path [edge] (init) edge node[anchor=west] {$x:=0$} (loop) ;
\path [edge] (loop) edge[loop left,distance=1cm] (loop) ;
\path [edge] (loop) edge node[anchor=east] {$x:=x+1$} (nondet) ;

\draw [transparent,use as bounding box](-3.3,0.4) rectangle(2.3,-3.2);

\path [edge] (nondet) edge [in=135,out=225,distance=30mm] node[anchor=east] {\parbox{12mm}{$x \geq 100$\\$x:=0$}} (loop) ;
\path [edge] (nondet) edge [in=45,out=325,distance=15mm] node[anchor=west] {$x < 100$} (loop) ;
\end{tikzpicture}}%
\hfill%
\subfigure[SSA version. $x=\phi(e_1,e_2,\dots)$ denotes a SSA $\phi$-node: $x$ takes value $e_1$ if control flows from the first incoming edge, $e_2$ from the second\dots]{\label{fig:listing1_ssa_cfg}%
\begin{tikzpicture}[node distance=1.2cm]
\node [loopblock] (init) {$p_1$} ;
\node [loopblock, below of=init] (loop) {$p_2: x_2 = \phi(x_1,x_2,x_3,x_4)$} ;
\node [block, below of=loop] (nondet) {$p_3$} ;
\path [edge] (init) edge node[anchor=west] {$x_1=0$} (loop) ;
\path [edge] (loop) edge[loop left,distance=12mm] (loop) ;
\path [edge] (loop) edge node[anchor=east] {$x_3:=x_2+1$} (nondet) ;

\draw [transparent,use as bounding box](-3.3,0.4) rectangle(2.3,-3.2);

\path [edge] (nondet) edge [in=160,out=225,distance=30mm] node[anchor=east] {\parbox{12mm}{$x_3 \geq 100$\\$x_4=0$}} (loop) ;
\path [edge] (nondet) edge [in=25,out=315,distance=30mm] node[anchor=east] {$x_3 < 100$} (loop) ;
\end{tikzpicture}}%
\caption{Control flow graph corresponding to listing~\ref{lst:circular}.}
\label{fig:cfg_for_lst_circular}
\end{figure}
We consider programs defined by a control flow graph: a set $P$ of control points, for each control point $p \in P$ a (possibly empty) set $I_p$ of initial values, a set $E \subseteq P \times P$ of directed edges, and the semantics $\tau_e: \parts{\Sigma} \rightarrow \parts{\Sigma}$ of each edge $e \in E$ where $\parts{\Sigma}$ is the set of possible values of the tuple of program variables. $\tau_e$ thus maps a set of states before the transition expressed by edge $e$ to the set of states after the transition.

To each control point $p \in P$ we attach a set $X_p \subseteq \Sigma$ of reachable values of the tuple of program variables at program point~$p$. The concrete semantics of the program is the least solution of a system of semantic equations~\cite{CousotCousot_JLC92}:
$X_p = I_p \cup \bigcup_{(p',p) \in E} \tau_{(p',p)} (X_{p'})$.

Abstract interpretation replaces the concrete sets of states in $\parts{\Sigma}$ by elements of an abstract domain $D$. In lieu of applying exact operations $\tau$ to sets of concrete program states, we apply abstract counterparts $\abstr{\tau}$.%
\footnote{Many presentations of abstract interpretation distinguish the abstract element $\abstr{x} \in D$ from the set of states $\gamma(\abstr{x})$ it represents. We opted not to, for the sake of brevity.}
An abstraction $\abstr{\tau}$ of a concrete operation $\tau$ is deemed to be correct if it never ``forgets'' states:
\begin{equation}\label{eqn:sound_transition}
\forall X \in D~  \tau(X) \subseteq \abstr{\tau}(X)
\end{equation}
We also assume an ``abstract union'' operation $\sqcup$, such that $X \cup Y \subseteq X \sqcup Y$.
For instance, $\Sigma$ can be $\bbQ^n$, $D$ can be the set of convex polyhedra and $\sqcup$ the convex hull operation~\cite{Halbwachs_PhD,CousotHalbwachs78,PPL}.

In order to find an inductive invariant, one solves a system of abstract semantic inequalities:
\begin{equation}\left\{
\label{eqn:inductive}
\begin{array}{l}
\forall p~ I_p \subseteq X_p\\
\forall (p',p) \in E~ \abstr{\tau}_{(p',p)} (X_{p'}) \subseteq X_p.\\
\end{array}
\right.\end{equation}
Since the $\abstr{\tau}_e$ are correct abstractions, it follows that any solution of such a system defines an inductive invariant; one wishes to obtain one that is as strong as possible (``strong'' meaning ``small with respect to $\subseteq$''), or at least sufficiently strong as to imply the desired properties.

Assuming that all functions $\abstr{\tau}_e$ are monotonic with respect to $\subseteq$,  and that $\sqcup$ is the least upper bound operation in $D$ with respect to $\subseteq$, one obtains a system of monotonic abstract equations:
$X_p = I_p \sqcup \bigsqcup_{(p',p) \in E} \abstr{\tau}_{(p',p)} (X_{p'})$.
If $(D,\subseteq)$ has no infinite ascending sequences ($d_1 \subsetneq d_2 \subsetneq \dots$ with $d_1, d_2, \dots \in D$), then one can solve such a system by iteratively replacing the contents of the variable on the left hand side by the value of the right hand side, until a fixed point is reached. The order in which equations are iterated does not change the final result.

Many interesting abstract domains, including that of convex polyhedra, have infinite ascending sequences. One then classically uses an extrapolation operator known as \emph{widening} and denoted by $\widening$ in order to enforce convergence within finite time. The iterations then follow the ``upward iteration scheme'':
\begin{equation}
X_p := X_p \widening \left(X_p \sqcup \bigsqcup_{(p',p) \in E} \abstr{\tau}_{(p',p)} (X_{p'}) \right)
\end{equation}
where the contents of the left hand side gets replaced by the value of the right hand side.
The convergence property is that any sequence $u_n$ of elements of $D$ of the form $u_{n+1} = u_n \widening v_n$, where $v_n$ is another sequence, is stationary~\cite{CousotCousot_JLC92}. It is sufficient to apply widening only at a set of program control nodes $P_W$ such that all cycles in the control flow graph are cut. Then, through a process of \emph{chaotic iterations} \cite[Def. 4.1.2.0.5, p.~127]{Cousot_PhD}, one converges within finite time to an inductive invariant satisfying Rel.~\ref{eqn:inductive}.

Once an inductive invariant is found, it is possible to improve it by iterating the $\abstr{\psi}$ function defined as $Y = \abstr{\psi}(X)$,
noting $X = (X_p)_{p \in P}$ and $Y = (Y_p)_{p \in P}$, with
$Y_p = I_p \sqcup \bigsqcup_{(p',p) \in E} \abstr{\tau}_{(p',p)} (X_{p'})$.
If $X$ is an inductive invariant, then for any $k$, ${\abstr{\psi}}^k(X)$ is also an invariant.
This technique is an instance of \emph{narrowing iterations}, which may help recover some of the imprecision induced by widening~\cite[\S4]{CousotCousot_JLC92}.

\begin{algorithm}
\caption{Classical Algorithm}
\label{algo:classic}
\begin{algorithmic}[1]
  \State $A\leftarrow \emptyset$;
  \ForAll{$p \in P$ such that $I_p\neq\emptyset$} 
  \State  $A\leftarrow A\cup\{p\}$ 
  \EndFor;\Comment Initialise $A$ to the set of
  all non empty initial nodes

  \While{$A$ is not empty} \label{step_while}\Comment Fixpoint Iteration

  \State Choose $p_1\in A$ \label{step_pick}
  \State  $A\leftarrow A\setminus\{p_1\}$ 

  \ForAll{outgoing edge (e) from $p_1$ }

  \State Let $p_2$ be the destination of $e$ :

  \If{$p_2\in P_W$}
  \State $X_{temp} \leftarrow X_{p_2}\widening \big(X_{p_2}\sqcup
  \tau_e^\sharp(X_{p_1})\big)$ \Comment Widening node;
  
  \Else 

  \State $X_{temp} \leftarrow X_{p_2}\sqcup \tau_e^\sharp(X_{p_1})$ ;

  \EndIf

  \If{$ X_{temp} \not\subseteq X_{p_2}$} \Comment The value must be updated
  \State $X_{p_2} \leftarrow X_{temp}$;
  \State $A\leftarrow A\cup\{p_2\}$;

  \EndIf
  \EndFor;

  \EndWhile; \Comment End of Iteration
  
  \State Possibly narrow
  
  \State \Return all $X_{p_i}$s; 
\end{algorithmic}
\end{algorithm}

\label{sec:naive}
A naive implementation of the upward iteration scheme described above
is to maintain a work-list of program points $p$ such that $X_p$ has
recently been updated and replaced by a strictly larger value (with
respect to $\subseteq$), pick and remove the foremost member $p$,
apply the corresponding rule $X_p := \dots$, and insert into the
work-list all $p'$ such that $(p,p') \in E$ (This algorithm is formally
described in Algorithm~\ref{algo:classic}).  

\paragraph{Example of Section~\ref{e:g} (Cont'd)}
Figure~\ref{fig:listing1_cfg} gives the control flow graph obtained by
compilation of Listing~\ref{lst:circular}. Node $p_2$ is the unique
widening node.

The classical algorithm (with the interval abstract domain) performs
on this control flow graph of  the
following iterations :
\begin{itemize}
\item Initialisation : $X_{p_1}\leftarrow (-\infty,+\infty)$,
  $X_{p_2}\leftarrow X_{p_3} \leftarrow X_{p_4}\leftarrow \emptyset$.
\item Step 1: $X_{p_2} \leftarrow [0,0]$, then the transition to
  $p_3$ is enabled, $X_{p_3} \leftarrow[1,1]$, then the return edge to
  $p_2$ gives the new point $x=1$ to $X_{p_2}$, the new polyhedron is
  then $X_{p_2}=[0,1]$ after performing the convex hull. Widening gives
  the polyhedron $X_{p_2}=[0,\infty)$.

  (The widening operator on
  intervals is defined as $[x_l,x_r]\widening[x'_l,x'_r]=[x"_l,x"_r]$
  where $x"_l=x_l$ if $x_l=x'_l$ else $-\infty$, and  $x"_r=x_r$ if
  $x_r=x'_r$ else $+\infty$.)

\item Step 2: $X_{p_3}$ becomes  $[1,+\infty)$. The second transition from
  $p_3$ to $p_2$ is thus enabled, and the back edge to $p_2$ gives the
  point $x=0$ to $X_{p_2}$. At the end of step 2 the convergence is
  reached.

\item If we perform a narrowing sequence, there is no gain of precision
  because of the simple loop over the control point $p_2$.
\end{itemize}


\section{Our Method}
\label{sec:method}

We have seen two examples of programs where classical polyhedral analysis fails to compute good invariants. How could we improve on these results?
\begin{itemize}
\item
In order to get rid of the imprecision in Listing~\ref{lst:sinc_minus1}, one could ``explode'' the
control-flow graph: in lieu of a sequence of $n$ if-then-else, with
$n$ merge nodes with $2$ input edges, one could distinguish the $2^n$
program paths, and having a single merge node with $2^n$ input
edges. As already pointed out, this would lead to exponential blowup in both time and space.
\item
One way to get rid of imprecision of classical analysis (Sec.~\ref{sec:background}) on the program from Fig.~\ref{fig:listing1_cfg} would be to consider each path through the loop at a time and compute a local invariant for this path. Again, the number of such paths could be exponential in the number of tests inside the loop.
\end{itemize}

The contribution of our article is a generic method that addresses both of these difficulties. 

\subsection{Reduced Transition Multigraph and Path Focusing}
\label{sec:transition_graph}

Consider a control flow graph $(P,E)$ with associated transitions
$(\tau_e)_{e \in E}$, a set of \emph{widening points} $P_W \subseteq
P$ such that removing $P_W$ cuts all cycles in the graph, and a set
$P_R$ of \emph{abstraction points}, such that $P_W \subseteq P_R
\subseteq P$ (On the figures, the nodes in $P_R$ are in bold). We make
no assumption regarding the choice of $P_W$; there are classical
methods for choosing widening
points~\cite[\S3.6]{Bourdoncle_PhD}. $P_R$~can be taken equal to
$P_W$, or may include other nodes; this makes sense only if these
nodes have several incoming edges. Including other nodes will tend to
reduce precision, but may improve scalability. We also make the
simplifying assumption that the set of initial values $I_p$ is empty
for all nodes in $P \setminus P_R$ --- in other words, the set of
possible control points at program start-up is included in~$P_R$.

We construct (virtually) the reduced control multigraph $(P_R,E_R)$, with edges $E_R$ consisting of the paths in $(P,E)$ that start and finish on nodes in $P_R$, with associated semantics the composition of the semantics of the original edges $\tau_{e_1 \rightarrow \dots \rightarrow e_n} = \tau_{e_n} \circ \dots \circ \tau_{e_1}$. There are only a finite number of such edges, because the original graph is finite and removing $P_R$ cuts all cycles. There may be several edges between two given nodes, because there may exist several control paths between these nodes in the original program.
Equivalently, this multigraph can be obtained by starting from the original graph $(P,E)$ and by removing all nodes $p$ in $P \setminus P_R$ as follows: each couple of edges $e_1$, from $p_1$ to $p$, and $e_2$, from $p$ to $p_2$, is replaced by a single edge from $p_1$ to $p_2$ with semantics $\tau_{p_2} \circ \tau_{p_1}$.
\paragraph{Example of Section~\ref{e:g} (Cont'd)}
The reduced control flow graph obtained for our running example is\vspace{-1em}
\begin{center}
\begin{tikzpicture}[node distance=1.2cm]
\node [loopblock] (init) {} ;
\node [loopblock, below of=init] (loop) {loop} ;
\path [edge] (init) edge node[anchor=west] {$x:=0$} (loop) ;
\path [edge] (loop) edge[loop below,distance=1cm] (loop) ;
\path [edge] (loop) edge [loop right,distance=1.5cm] node {\parbox{2.5cm}{guard $x \geq 99$\\ $x:=0$}} (loop) ;
\path [edge] (loop) edge [loop left,distance=1.5cm] node {\parbox{2.5cm}{guard $x < 99$\\ $x:=x+1$}} (loop) ;
\end{tikzpicture}
\end{center}\vspace{-1em}

Our analysis algorithm performs chaotic iterations over that reduced multigraph, without ever constructing it explicitly. We start from an iteration strategy, that is, a method for choosing which of the equations to apply next; one may for instance take a variant of the naive ``breadth-first'' algorithm from \S\ref{sec:naive}, but any iteration strategy~\cite[\S3.7]{Bourdoncle_PhD} befits us (see also Alg.~\ref{algo:classic}). An iteration strategy maintains a set of ``active nodes'', which initially contains all nodes $p$ such that $I_p \neq \emptyset$. It picks one edge $e$ from an active node $p_1$ to a node $p_2$, and applies $X_{p_2} := X_{p_2} \sqcup \abstr{\tau}_e (X_{p_1})$ in the case of a node $p_2 \in P_R \setminus P_W$, and applies $X_{p_2} := X_{p_2} \widening (X_{p_2} \sqcup \abstr{\tau}_e (X_{p_1}))$ if $p_2 \in P_W$; then $p_2$ is added to the set of active nodes if the value of $X_{p_2}$ has changed.

Our alteration to this algorithm is that we only pick edges $e$ from $p_1$ to $p_2$ such that there exist $x_1 \in X_{p_1}$, $x_2 \in \tau_e(\{x_1\})$ and $x_2 \notin X_{p_2}$ with the current values of $X_{p_1}$ and $X_{p_2}$. In other words, going back to the original control flow graph, we only pick paths that add new reachable states to their end node, and we temporarily \emph{focus} on such a path.

How do we find such edges $e$ out of potentially exponentially many? We express them as the solution of a \emph{bounded reachability} problem --- how can we go from control state $p_1$ with variable state in $X_{p_1}$ to control state $p_2$ with variable state in $X_{p_2}$ ---, which we solve using satisfiability modulo theory (SMT). (See Alg.~\ref{algo:smtfixpoint})

\subsection{Finding Focus Paths}

\label{sec:focus_path}
We now make the assumption
that both the program transition semantics $\tau_e$ and the abstract elements $\abstr{x} \in D$ can be expressed within a decidable theory~$T$
(this assumption may be relaxed by replacing the concrete semantics,
including e.g. multiplicative arithmetic, by a more abstract one through
e.g. linearization~\cite{mine:vmcai06}).

Such is for instance the case if the program operates on rational
values, so a program state is an element of $\Sigma = \bbQ^n$, all
operations in the program, including guards and assignments, are
linear arithmetic, and the abstract domain is the domain of convex
polyhedra over $\bbQ^n$, in which case $T$ can be the theory of linear
real arithmetic (LRA). If program variables are integer, with program
state space $\Sigma = \bbZ^n$, but still retaining the abstract domain
of convex polyhedra over $\bbQ^n$, then we can take $T$ to be the
theory of linear integer arithmetic (LIA). Deciding the satisfiability
of quantifier-free formulas in either LIA or LRA, with atoms
consisting in propositional variables and in linear (in)equalities
with integer coefficients, is NP-complete.
There however exist
efficient decision procedures for such formulas, known as SMT-solvers,
as well as standardised theories and file formats~\cite{SMTLIB};
notable examples of SMT-solvers capable of dealing with LIA and LRA
are Z3 and Yices.
Kroening \& Strichman \cite{Kroening_Strichman_08} give a good introduction to the
techniques and algorithms in SMT solvers.

We assume that the program is expressed in SSA form, with each program variable being assigned a value at only a single point within the program~\cite{Cytron_et_al_POPL89}; standard techniques exist for converting to SSA. Figure~\ref{fig:cfg_for_lst_circular} gives both ``normal'' and SSA-form control-flow graphs for Listing~\ref{lst:circular}.

We transform the original control flow graph $(P,E)$ in SSA form by
disconnecting the nodes in $P_R$: each node $p_r$ in $P_R$ is split
into a ``source'' node $p^s_r$ with only outbound edges, and a
``destination'' node $p^d_r$ with only inbound edges. We call the
resulting graph
$(P',E')$. Figure~\ref{fig:listing1_disconnected_cfg_left} gives the
disconnected SSA form graph for Listing~\ref{lst:circular} where
$p_1$ and $p_2$ have been split.

We consider execution traces starting from a $p^s_r$ node and ending
in a $p^d_r$ node. We define them as for doing bounded model
checking~\cite{Armando_et_al_STTT09}.  To each node $p \in P'$ we
attach a Boolean~$b_p$ or \emph{reachability
  predicate}, expressing that the trace goes through program
point~$p$. For nodes $p'$ not of the form $p^s_r$, we have a constraint
$b_{p'} = \bigvee_p e_{p,p'}$, for $e_{p,p'}$ ranging over all incoming edges.
To each edge $p \rightarrow p'$ we attach a Boolean $e_{p,p'}$, and a
constraint $e_{p,p'} = b_p \land \tau_{p,p'}$.
The conjunction $\rho$ of all these constraints,
expresses the transition relation
between the $p^s_r$ and $p^d_r$ nodes
(with implicit existential quantification).

If the transitions $\tau_{(p,p')}$ are non-deterministic, a little care must be exercised for the path obtained from the $b_p$ to be unique. For instance, if from program point $p_1$ one can move non-deterministically to $p_2$ or $p_3$ through edges $e_2$ and $e_3$ an incorrect way of writing the formula would be $(b_2=e_2) \land (b_3=e_3) \land (e_2=b_1) \land (e_3=b_1)$, in which case $b_2$ and $b_3$ could be simultaneously true. Instead, we introduce special ``choice'' variables $c_i$ that model non-deterministic choices (Fig.~\ref{fig:listing1_focused_cfg}).

\begin{figure}[htb]
\subfigure[Disconnected (SSA) CFG]{\label{fig:listing1_disconnected_cfg_left}%
\begin{tikzpicture}[node distance=1.5cm]
\node [loopblock] (init) {$p_1^s$} ;
\node [loopblock, node distance=20mm, right of=init] (loops) {$p_2^s$} ;
\node [block, below of=loops] (nondet) {$p_3$} ;
\node [loopblock, below of=nondet, node distance=20mm] (loopd) {$p_2^d$ : $x'_2=\phi(x_1,x_4,x_3,x_2)$} ;
\path [edge] (loops) edge node[edgelabel] {$e_3$} node[left=1ex] { $x_3 = x_2+1$ } (nondet) ;
\path [edge] (nondet) edge node[edgelabel] {$e_5$} node[anchor=east] {\parbox{12mm}{$x_3 \geq 100$\\$x_4=0$}} (loopd) ;
\path [edge] (nondet) edge[out=315,in=45]  node[edgelabel] {$e_4$} node[right=1ex] {$x_3 < 100$} (loopd) ;
\path [edge] (loops) edge[out=330,in=0] node[edgelabel] {$e_2$} (loopd) ;
\path [edge] (init) edge[out=270,in=180] node[edgelabel] {$e_1$} node[left=1ex] {$x_1 = 0$} (loopd);
\end{tikzpicture}\hspace*{-9mm}}%
\hfill
\subfigure[With a focus path (solid edges) from $x_2=0$ at program point $2$ to $x'_2=1$ at the same program point]{\label{fig:listing1_disconnected_cfg}%
\begin{tikzpicture}[node distance=1.5cm]
\node [dashedblock] (init) {$p_1^s$} ;
\node [loopblock, node distance=20mm, right of=init] (loops) {$p_2^s$} ;
\node [block, below of=loops] (nondet) {$p_3$} ;
\node [loopblock, below of=nondet, node distance=20mm] (loopd) {$p_2^d$ : $x'_2=\phi(x_1,x_4,x_3,x_2)$} ;
\path [edge] (loops) edge node[edgelabel] {$e_3$} node[left=1ex] { $x_3 = x_2+1$ } (nondet) ;
\path [dashededge] (nondet) edge node[edgelabel] {$e_5$} node[anchor=east] {\parbox{12mm}{$x_3 \geq 100$\\$x_4=0$}} (loopd) ;
\path [edge] (nondet) edge[out=315,in=45] node[edgelabel] {$e_4$} node[right=1ex] {$x_3 < 100$} (loopd) ;
\path [dashededge] (loops) edge[out=330,in=0] node[edgelabel] {$e_2$}  (loopd) ;
\path [dashededge] (init) edge[out=270,in=180] node[edgelabel] {$e_1$} node[left=1ex] {$x_1 = 0$} (loopd);
\end{tikzpicture}\hspace*{-9mm}}

$(e_1 = (x_1 = 0) \land b_1^s) \allowbreak\land\allowbreak
 (e_3 = (x_3 = x_2+1) \land b_2^s \land c_2^s) \allowbreak\land\allowbreak
 (e_2 = b_2^s \land \neg c_2^s) \allowbreak\land\allowbreak
 (e_5 = b_3 \land x_3 \geq 100 \land x_4 = 0) \allowbreak\land\allowbreak
 (e_4 = b_3 \land x_3 < 100) \allowbreak\land\allowbreak
 (b_3 = e_3) \allowbreak\land\allowbreak
 (b_2^d = e_1 \lor e_4 \lor e_5 \lor e_2) \allowbreak\land\allowbreak
 (x'_2 = \ite{e_1}{x_1}{\ite{e_5}{x_4}{\ite{e_4}{x_3}{x_2}}})$

\caption{Disconnected version of the SSA control flow graph of Fig.~\ref{fig:listing1_ssa_cfg}, and the corresponding SMT formula. $\ite{b}{e_1}{e_2}$ is a SMT construct whose value is ``if $b$ then the value of $e_1$ else the value of $e_2$''. To each node $p_x$ corresponds a Boolean $b_x$ and an optional choice variable~$c_x$; to each edge, a Boolean $e_y$.}
\label{fig:listing1_focused_cfg}
\end{figure}

In order to find a path from program point $p_1 \in P_R$, with variable state $x_1$, to program point $p_2 \in P_R$, with variable state $x_2$, we simply conjoin $\rho$ with the formulas $x_1 \in X_{p_1}$ and $x_2 \notin X_{p_2}$, with $x_1$, $x_2$, $x_1 \in X_{p_1}$ and $x_2 \notin X_{p_2}$ expressed in terms of the SSA variables.%
\footnote{The formula defining the set of values represented by an abstract element $X$ has sometimes been denoted by $\hat{\gamma}$~\cite{DBLP:conf/vmcai/RepsSY04}.}
For instance, if $X_{p_1}$ and $X_{p_2}$ are convex polyhedra defined by systems of linear inequalities, one simply writes these inequalities using the names of the SSA-variables at program points $p_1$ and~$p_2$.

We apply SMT-solving over that formula. The result is either ``unsatisfiable'', in which case there is no path from $p_1$, with variable values $x_1$, to $p_2$, with variable values $x_2$, such that $x_1 \in X_{p_1}$ and $x_2 \notin X_{p_2}$, or ``satisfiable'', in which case SMT-solving also provides a model of the formula (a satisfying assignment of its free variables); from this model we easily obtain such a path, unique by construction of~$\rho$.

Indeed, a model of this formula yields a trace of
execution: those $b_p$ predicates that are true designate the program
points through which the trace goes, and the other variables give the
values of the program variables.

\paragraph{Example of Section~\ref{e:g} (Cont'd)}
The SSA form of the control flow graph of
Figure~\ref{fig:listing1_cfg} is depicted in
Figure~\ref{fig:listing1_ssa_cfg}.
Fig.~\ref{fig:listing1_focused_cfg} shows the disconnected version of
the SSA Graph (the node $p_2$ is now split), and the formula $\rho$
expressing the semantics is shown beneath it.

Then, consider the problem of finding a path starting in control point
$2$ inside polyhedron $x=0$ and ending at the same control point but
outside of that polyhedron.  Note that because there are two outgoing
transitions from node $p_2^s$, which are chosen non-deterministically,
we had to introduce a Boolean choice variable~$c_2^s$.

The focus path of Fig.~\ref{fig:listing1_disconnected_cfg} was
obtained by solving the formula $\rho \land b_1^s=\false \land
b_2^s=\true \land b_2^d=\true \land (x_2=0) \land \neg(x'_2=0)$: we
impose that the path starts at point $p_2^s$ (thus forcing
$b_1^s=\false \land b_2^s=\true$) in the polyhedron $x=0$ (thus
$x_2=0$) and ends at point $p_2^d$ (thus forcing $b_2^p=\true$)
outside of that polyhedron (thus $\neg(x_2=0)$).

\subsection{Algorithm}
\label{sec:algorithm}

Algorithm~\ref{algo:smtfixpoint} consists in the iteration of the path finding method of Sec.~\ref{sec:focus_path}, coupled with forward abstract interpretation along the paths found and, optionally, path acceleration.

\begin{algorithm}
\caption{Path-focused Algorithm}
\label{algo:smtfixpoint}
\begin{algorithmic}[1]

  \State Compute SSA-form of the control flow graph.

  \State Choose $P_R$, compute the disconnected graph $(P',E')$ accordingly.

  \State $\rho\leftarrow \mbox{computeFormula}(P',E')$ 
  \Comment Precomputations

  \State $A\leftarrow \emptyset$;
  \ForAll{$p \in P_R$ such that $I_p\neq\emptyset$} 
  \State  $A\leftarrow A\cup\{p\}$ 
  \EndFor;

  \While{$A$ is not empty} \Comment Fixpoint Iteration on the reduced graph

  \State Choose $p_1\in A$ 
  \State  $A\leftarrow A\setminus\{p_1\}$ 

  \Repeat
  \State $\displaystyle res \leftarrow \mbox{SmtSolve}
   \left( \rho \wedge b_{p_1} \land x_1\in X_{p_1}
  \land \bigvee_{p_2 \mid (p_1,p_2) \in E'}
               \left(b_{p_2} \land x_2\not\in X_{p_2}\right) \right)$

  \If{$res$ is not ``unsat''}
  \State Compute $e'\in E'$ from $res$
  \Comment Extraction of path from the model (\S\ref{sec:focus_path})
  \State $Y\leftarrow \tau_{e'}^\sharp(X_{p_1})$

  \If{$p_2\in P_W$}
   \State $X_{temp} \leftarrow X_{p_2}\widening \big(X_{p_2}\sqcup Y\big)$ 
     \Comment Final point $p_2$ is a widening point
  \Else 

  \State $X_{temp} \leftarrow X_{p_2}\sqcup Y$

  \EndIf

  \Comment at this point
    $X_{temp} \not\subseteq X_{p_2}$ otherwise $p_2$ would not have
    been chosen
    \label{lab:comment_y_greater}
  \State $X_{p_2} \leftarrow X_{temp}$
  \State $A\leftarrow A\cup\{p_2\}$

   \EndIf
  \Until $res$=``unsat'' 

  \EndWhile \Comment End of Iteration
  \State Possibly narrow (see Sec.~\ref{sec:narrowing})
  \State Compute $X_{p_i}$ for $p_i\not\in P_R$
  \State \Return all $X_{p_i}$
\end{algorithmic}
\end{algorithm}

\subsection{Correctness and Termination}
\label{sec:correctness}
We shall now prove that this algorithm terminates, and that the resulting $X_p$ define an inductive invariant that contains all initial states~$I_p$. The proof is a variant of the correctness proof of the chaotic iterations.

The invariant maintained by this algorithm is that all nodes $p_1 \in P_R \setminus A$ are such that there is no execution trace starting at point $p_1$ in a state $x_1 \in X_{p_1}$ and ending at point $p_2$ in a state $x_2 \notin X_{p_2}$. Evidently, if $A$ becomes empty, then this condition means that $X_p$ is an inductive invariant.

Termination is ensured by the classical argument of termination of chaotic iterations in the presence of widening: they always terminate if all cycles in the control flow graph are broken by widening points~\cite[Th.~4.1.2.0.6, p.~128]{Cousot_PhD}. In short, an infinite iteration sequence is bound to select at least one node $p$ in $P_W$ an infinite amount of times, because $P_W$ breaks all cycles, but due to the properties of widening, $X_p$ should be stationary, which contradicts the infinite number of selections.
Our comment at line \ref{lab:comment_y_greater} of Alg.~\ref{algo:smtfixpoint} is important for termination: it ensures that for any widening node $p$, the sequence of values taken by $X_p$ when it is updated and reinserted into set $A$ is strictly ascending, which ensures termination in finite time.

\subsection{Self-Loops}
\label{sec:self-loops}
The algorithm in the preceding subsection is merely a ``clever'' implementation of standard polyhedral analysis \cite{CousotHalbwachs78,Halbwachs_PhD} on the reduced control multigraph $(P_R,E_R)$; the difference with a naive implementation is that we do not have to explicitly enumerate an exponential number of paths and instead leave the choice of the focus path to the SMT-solver. We shall now describe an improvement in the case of self-loops, that is, single paths from one node to itself.

\begin{algorithm}
\caption{Path-focused Algorithm with Self-Loops. \stressalgo{~} marks changes from Alg.~\ref{algo:smtfixpoint}.}
\label{algo:smtfixpoint_selfloops}
\begin{algorithmic}[1]

  \State Compute SSA-form of the control flow graph.

  \State Choose $P_R$, compute the disconnected graph $(P',E')$ accordingly.

  \State $\rho\leftarrow \mbox{computeFormula}(P',E')$ 
  \Comment Precomputations

  \State $A\leftarrow \emptyset$;
  \ForAll{$p \in P_R$ such that $I_p\neq\emptyset$} 
  \State  $A\leftarrow A\cup\{p\}$ 
  \EndFor;

  \While{$A$ is not empty} \Comment Fixpoint Iteration on the reduced graph

  \State Choose $p_1\in A$ 
  \State  $A\leftarrow A\setminus\{p_1\}$
  \State  \stressalgo{$U = \emptyset$}
    \Comment $U$ is a set of ``already seen'' edges
  \Repeat
    \State $\displaystyle res \leftarrow \mbox{SmtSolve}
            \left( \rho \wedge b_{p_1} \land x_1\in X_{p_1}
              \land \bigvee_{p_2 \mid (p_1,p_2) \in E'}
               \left(b_{p_2} \land x_2\not\in X_{p_2}\right) \right)$

    \If{$res$ is not ``unsat''}
    \State Compute $e'\in E'$ from $res$
    \If{\stressalgo{$p_1 = p_2$}}
      \State \stressalgo{$Y\leftarrow \textit{loopiter}(\tau_{e'}^\sharp,X_{p_1})$}
    \Else
      \State $Y\leftarrow \tau_{e'}^\sharp(X_{p_1})$
    \EndIf

    \If{$p_2\in P_W$ \stressalgo{$\mbox{\bf and } (p_1 \neq p_2 \lor e' \in U)$}}
     \State $X_{p_2} \leftarrow X_{p_2}\widening \big(X_{p_2}\sqcup Y\big)$ 
       \Comment Final point $p_2$ is a widening point
    \Else 
       \State $X_{p_2} \leftarrow X_{p_2}\sqcup Y$
       \State \stressalgo{$U \leftarrow U \cup \{ e' \}$}
    \EndIf

        \State $A\leftarrow A\cup\{p_2\}$
      \EndIf
    \Until $res$=``unsat'' 

  \EndWhile \Comment End of Iteration
  \State Compute $X_{p_i}$s for $p_i\not\in P_R$
  \State \Return all $X_{p_i}$s
\end{algorithmic}
\end{algorithm}

Algorithm \ref{algo:smtfixpoint_selfloops} is a variant of Alg.~\ref{algo:smtfixpoint} where self-loops are treated specially:
\begin{itemize}
\item The $\textit{loopiter}(\abstr{\tau},X)$ function returns the result of a widening / narrowing iteration sequence for abstract transformer $\abstr{\tau}$ starting in $X$; it returns $X'$ such that $X \subseteq X'$ and $\abstr{\tau}(X') \subseteq X'$.

\item In order not to waste the precision gained by \textit{loopiter}, the first time we consider a self-loop $e'$ we apply a union operation instead of a widening; set $U$ records the self-loops that have already been visited. This is a form of delayed widening~\cite{Halbwachs_CAV93}.
\end{itemize}

Termination is still guaranteed, because the inner loop cannot loop forever: it can visit any self-loop edge $e'$ at most once before applying widening.

\newpage
\paragraph{Example of Section~\ref{e:g} (Cont'd)}
Let us perform our algorithm on our example~:
\begin{itemize}
\item Step 1 : Is there a path from control point $p_1$ to control
  point $p_2$ feasible (without additional constraint) ? Yes. On
  Figure~\ref{fig:listing1_focused_cfg}, the obtained model
  corresponds to the transition from $p_1^s$ to $p_2^d$, and leads to
  the interval $X_{p_2}=[0,0]$.

\item Step 2 : Is there a path from $p_2$ with $x=0$ to $p_2$ with
  $x\neq 0$ ? The answer to this query is depicted in
  Figure~\ref{fig:listing1_disconnected_cfg}: there is such a path,
  on which we now focus. This path is considered as a loop
  and we therefore do a local iteration with widenings (\textit{loopiter}).
  $X_{p_2}$ becomes $[0,1]$, then after
  widening $[0,\infty]$. A narrowing step gives finally
  $X_{p_2}=[0,99]$, which is thus the result of \textit{loopiter}.

\item Step 3 : Is there a path from $p_2$ with $x\in [0,99]$ to
  $p_2$ with $x' \notin [0,99]$ ? No.
\end{itemize}
The iteration thus ends with the desired invariant.


\section{Extensions}
\label{sec:extensions}
\subsection{Narrowing}
\label{sec:narrowing}
Narrowing iterations can also be applied within our framework. Let us assume that some inductive invariant $X_{p \in P_R}$ has been computed; it satisfies the relation $\psi(X) \subseteq X$ component-wise, noting $X = (X_1, \dots, X_{|P|})$,
and $\psi(X)$ denotes $(Y_1, \dots, Y_{|P|})$ defined as
\begin{equation}
Y_{p_2} = I_{p_2} \cup
  \bigcup_{e \in E_R \text{~$e$ from $p_1$ to $p_2$}} \tau_e\left(X_{p_1}\right)
\end{equation}
The abstract counterpart to this operator is $\abstr{\psi}$, defined similarly, replacing $\tau$ by $\abstr{\tau}$ and $\cup$ by~$\sqcup$.
It satisfies the correctness condition (see Rel.~\ref{eqn:sound_transition})
$\forall X \in D~ \psi(X) \subseteq \abstr{\psi}(X)$.

As per the usual narrowing iterations, we compute a narrowing sequence $X^{(k)} = {\abstr{\psi}}^k(X)$. It is often sufficient to stop at~$k=1$; otherwise one may stop when $X^{(k+1)} \nsubseteq X^{(k)}$. Let us now see a practical algorithm for computing $Y = \abstr{\psi}(X)$:

For all $p \in P_R$, we initialise $Y_p := I_p$. 
For all $p_2 \in P_R$, we consider all paths $e \in E_R$ from $p_1 \in P_R$ to $p_2$ such that there exist $x_1 \in X_{p_1}$, $x_2 \in X_{p_2}$, $x_2 \in \tau_e(\{x_1\})$ as explained in \S\ref{sec:focus_path}. We then update $Y_{p_2} := Y_{p_2} \sqcup \abstr{\tau}_e(X_{p_1})$.

\subsection{Acceleration}
\label{subsec:accel}
In Sec.~\ref{sec:self-loops}, we have described \textit{loopiter} function that performs a classical widening / narrowing iteration over a single path. In fact, the only requirement over it is that $\textit{loopiter}(\abstr{\tau},X)$ returns $X'$ such that $X \subseteq X'$ and $\abstr{\tau}(X') \subseteq X'$. In other words, $X'$ is an over-approximation of ${\abstr{\tau}}^*(X)$, noting $R^*$ the transitive closure of~$R$.

In some cases, we can compute directly such an over-approximation, sometimes even obtaining ${\abstr{\tau}}^*(X)$ exactly; this is known as \emph{acceleration} of the loop. Examples of possible accelerations include the case where $\tau_e$ is given by a difference bound matrix \cite{DBLP:conf/cav/ComonJ98}, an octagon \cite{VERIMAG_TR-2008-16}, ultimately periodic integer relations \cite{Bozga_et_al_TR-2010-3} or certain affine linear relations~\cite{DBLP:conf/sas/GonnordH06,Gonnord_PhD,DBLP:journals/entcs/AncourtCI10}.

For instance, the focus path of Fig.~\ref{fig:listing1_disconnected_cfg} consists in the operations and guards $x=x+1; x < 100$; instead of iterating that path, we can compute its exact acceleration, yielding $x \in [0,99]$.

\subsection{Partitioning}
\label{sec:partitioning}
It is possible to partition the states at a given program point according to some predicate or a partial history of the computation~\cite{Rival_Mauborgne_TOPLAS07}. This amounts to introducing several graph nodes representing the same program point, and altering the transition relation.

\subsection{Input-Output Relations}
\label{sec:input-output}
As with other analyses using relational domains, it is possible to obtain abstractions of the input-output relation of a program block or procedure instead of an abstraction of the set of states at the current point \cite{DBLP:journals/entcs/AncourtCI10}; this also allows analyzing recursive procedures~\cite[Sec.~7.2]{Halbwachs_PhD}. It suffices to include in the set of variables copies of the variables at the beginning of the block or procedure; then the abstract value obtained at the end of the block or procedure is the desired abstraction.

\section{Implementation and Preliminary Results} \label{experiments}

Our algorithm has been implemented as an option for \aspic, that
computes invariants from counter automata with Linear Relation
Analysis (\cite{feautriergonnord_tapas2010}). We wrote an Ocaml
interface to the Yices SMT-solver (\cite{DBLP:conf/cav/DutertreM06}),
and modified the  fixpoint computation inside \aspic to deal with
local iterations of paths. The implementation still needs some
improvements, but the preliminary results are promising, and we
describe some of them in Table~\ref{theridiculousbenchs}. We provide
no timing results since we were unable to detect any overcost due to
the method. These two examples show that since we avoid (some) convex
hulls, the precision of the whole analysis is improved.
\begin{table}
\caption{Invariant generation on two simple challenging programs}
\begin{tabular}{|c|c|c|}
\hline
Program  & Automaton &  Result and notes\\
\hline 
\begin{minipage}[c]{0.40\linewidth}
\begin{lstlisting}[caption={Boustrophedon},label=lst:boustrophedon]
void boustrophedon() {
  int x;
  int d;
  x = 0;
  d = 1;
  while (1) {
    if (x == 0) d=1;
    if (x == 1000) d=-1;
    x += d;
  }
}
\end{lstlisting}
\end{minipage}&
\begin{minipage}[c]{0.30\linewidth}
\includegraphics[scale=0.25]{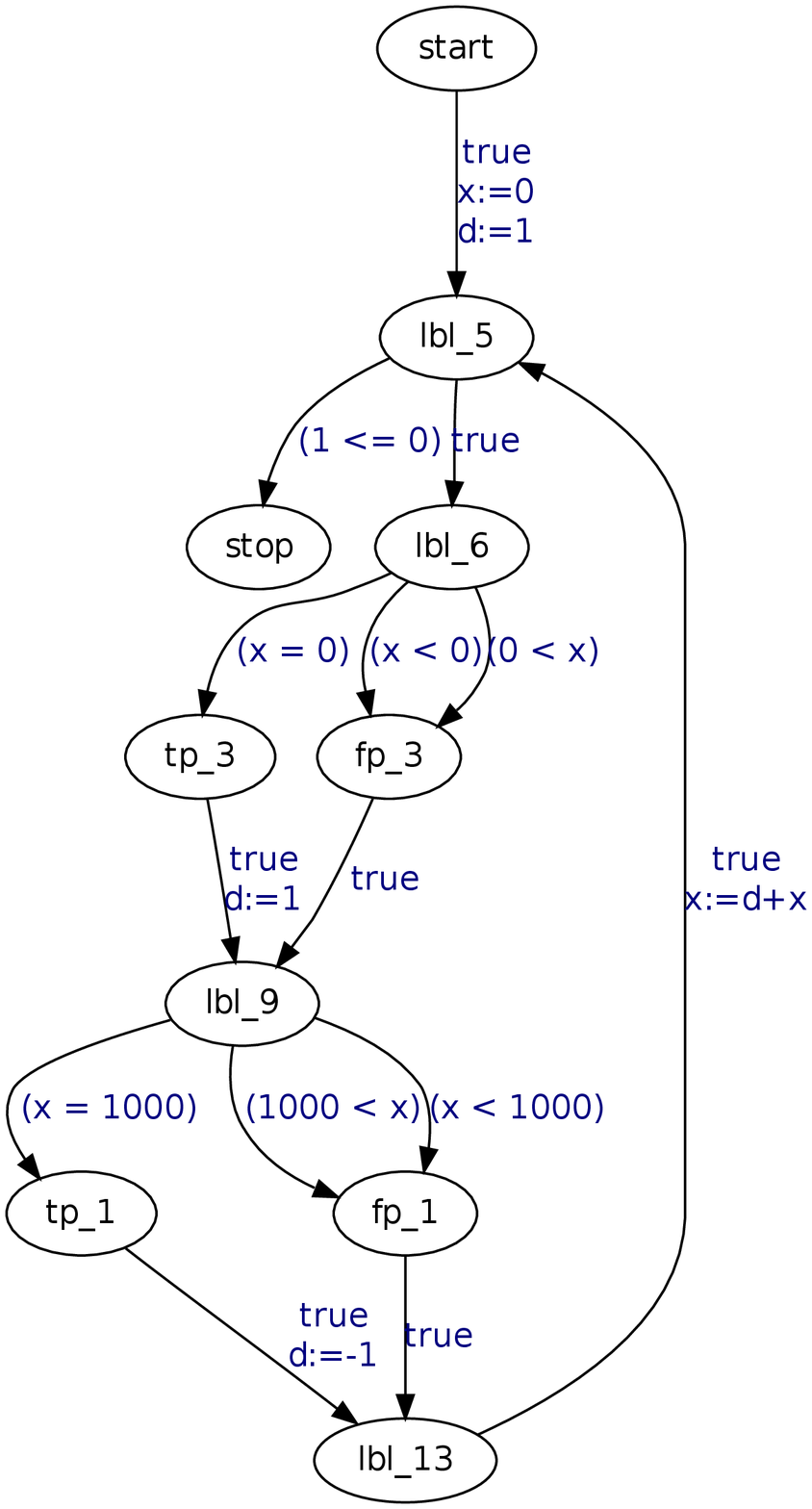}
\end{minipage}&
 \begin{minipage}[c]{0.30\linewidth}
   The compilation of the program gives an expanded control structure
   where some paths are ``clearly'' unfeasible (e.g. imposing both $x
   < 0$ and $x > 1000$), thus the only feasible ones are guarded by $x
   < 0$, $x = 0$, $0 < x < 1000$, $x = 1000$ and $x > 1000$.

   The tool finds the invariant $\{\mathbf{0 \leq  x \leq 1000, -1\leq
     d\leq 1}\}$\\
   Classical Analysis with widening ``upto'' gives $\{d\leq 1, d+1999\geq
   2x\}$ and Gopan and Reps' improvement is not able to find $x\geq 0$.
\end{minipage}
\\
\hline 
\begin{minipage}[c]{0.40\linewidth}
\begin{lstlisting}[caption={Rate limiter},label=lst:rlim]
void main() {
  float x_old, x;
  x_old = 0;
  while (1) {
    x = input(-1000,1000);
    if (x >= x_old+1)
        x = x_old+1;
    if (x <= x_old-1)
        x = x_old-1;
    x_old = x;
  }
}
\end{lstlisting}
Source : \cite{Monniaux_LMCS10}
\end{minipage}&
\begin{minipage}[c]{0.30\linewidth}
\includegraphics[scale=0.25]{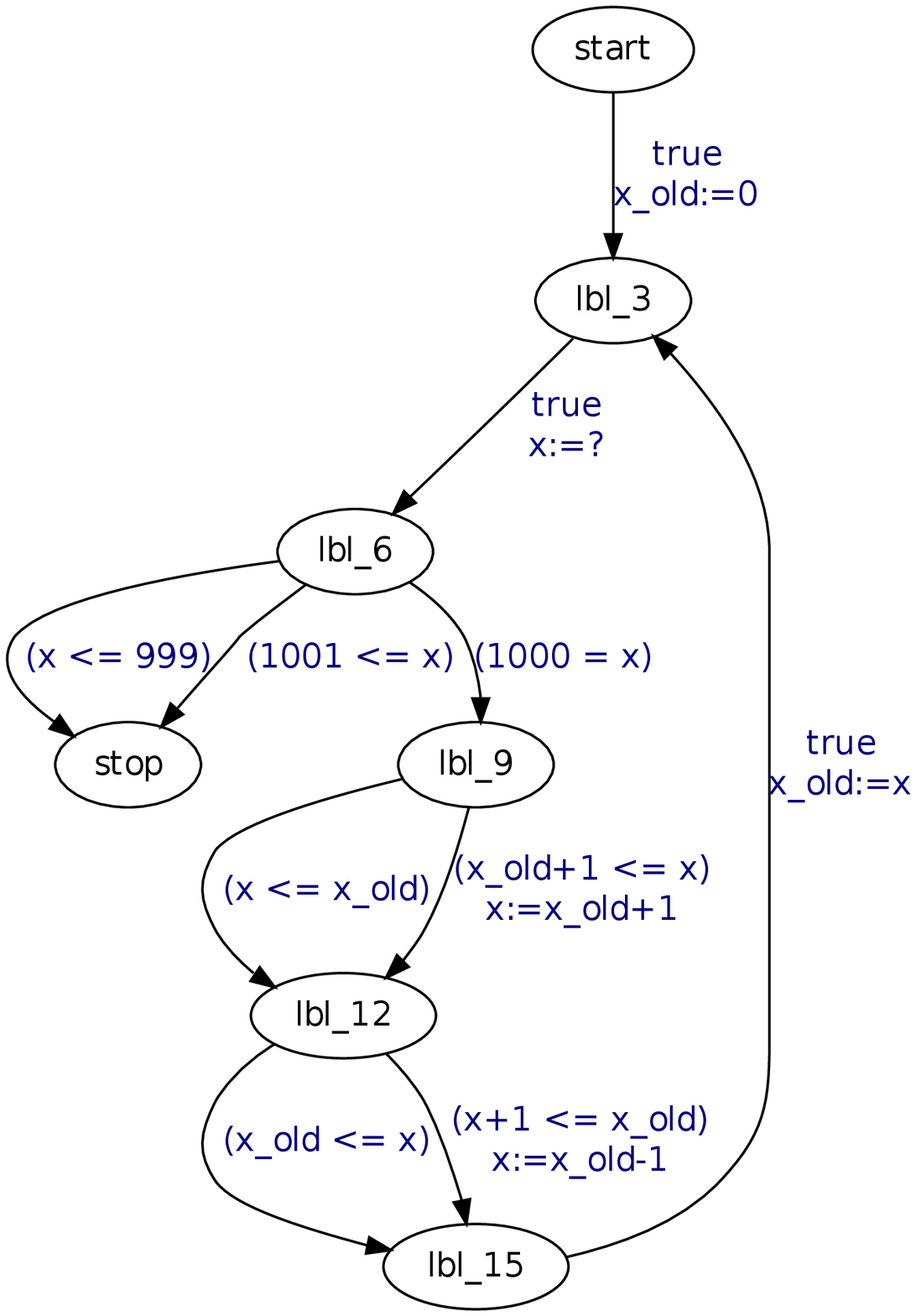}
\end{minipage}&
 \begin{minipage}[c]{0.30\linewidth}
In order to
properly analyse such a program, \textsc{Astr\'ee} distinguishes all
four execution paths inside the loop through \emph{trace
  partitioning}~\cite{Rival_Mauborgne_TOPLAS07}, which is triggered by
ad hoc syntactic criteria (e.g. two successive if-then-else). 
Our algorithm finds the invariant $\{\mathbf{-1000 \leq  x_{old} \leq
  1000}\}$, which is not found by classical analysis.
\end{minipage} \\
\hline
\end{tabular}
\label{theridiculousbenchs}
\end{table}

The rate limiter example is particularly interesting, since, like the
one in Listing~\ref{lst:sinc_minus1} (which does not include a loop),
it will be imprecisely analyzed by any method enforcing convex invariants
at intermediate steps.

\section{Related Work}

Our algorithm may be understood as a form of \emph{chaotic iterations}
\cite[\S2.9.1, p.~53]{Cousot_PhD} over a certain system of semantic
questions; we use SMT as an oracle to know which equations need
propagating.  The choice of widening points, and the order in which to
solve the abstract equations, have an impact on the precision of the
whole analysis, as well as its running time. Even though there exist
few hard general results as to which strategy is best \cite[\S4.1.2,
p.~125]{Cousot_PhD}, some methods tend to experimentally behave
better~\cite{Bourdoncle_PhD}.

``Lookahead widening'' \cite{DBLP:conf/cav/GopanR06} was our main
source of inspiration: iterations and widenings are adapted according
to the discovery of new feasible paths in the program. This approach
avoids loss of precision due to widening in programs with multiple
paths inside loops. It has proved its efficacy to suppress some
gross over-approximations induced by naive widening. However, it does
not solve the imprecisions introduced by convex hull (e.g. it produces
false alarms on Listing~\ref{lst:sinc_minus1}).

Our method analyzes separately the paths between cut-nodes.
We have pointed out that this is (almost) equivalent to considering finite
unions of elements of the abstract domain, known as the \emph{finite
  powerset} construction, between the cut-nodes.%
\footnote{It is equivalent if the only source of disjunctions are the
  splits in the control flow, and not atomic operations. For instance,
  if the test $|x| \geq 1$ is considered an atomic operation, then we
  could take the disjunction $x \geq 1 \lor x \leq -1$ as output. We
  can rephrase that as a control flow problem by adding a test $x \geq
  0$, otherwise said to express $|x|$ as a piecewise linear function
  with explicit tests for splits between the pieces.}  The finite
powerset construction is however costly even for loop-free code, and
it is not so easy to come up with widening operators to apply it to
codes with loops or recursive
functions~\cite{DBLP:journals/sttt/BagnaraHZ07}; for limiting the number
of elements in the unions, some may be lumped together (thus generally
introducing further over-approximation) according to affinity heuristics
\cite{Sriram_et_al_SAS06,Popeea:2006:IDP:1782734.1782760}.

Still, in the recent years, much effort has been put into the discovery of \emph{disjunctive invariants}, for instance in predicate abstraction
\cite{Gulwani_et_al_VMCAI09}. 
Of particular note is the recent work by Gulwani and Zuleger on inferring
disjunctive invariants \cite{DBLP:conf/pldi/GulwaniZ10} for finding bounds
on the number of iterations of loops. We improve on their method on two
points:
\begin{itemize}
\item In contrast to us, they assume that the transition
relation is given in disjunctive normal form \cite[Def.~5]{DBLP:conf/pldi/GulwaniZ10}, which in general has exponential size in the number of tests inside the loop. By using SMT-solving, we keep the DNF implicit and thus avoid this blowup.
\item By using acceleration, we may obtain more precise results than using widening, as they do for lattices that do not satisfy the ascending chain condition.
\end{itemize}

Nevertheless, their method allows expressing disjunctive invariants at
loop heads, and not only at intermediate points, as we do. However,
we think it is possible to get the best of both worlds and combine our method
with theirs. In order to obtain a disjunctive
invariant, they first choose a ``convexity witness'' (given that the number
of possible witnesses is exponential, they choose it using heuristics) \cite[p.~7]{DBLP:conf/pldi/GulwaniZ10}, and then they compute a ``transitive closure'' \cite[Fig.~6]{DBLP:conf/pldi/GulwaniZ10}, which is a form of fixed point iteration of input-output relations (as in our Sec.~\ref{sec:input-output}) over an expanded control-flow graph. The choice of the convexity witness amounts to a partitioning of the nodes and transition (Sec.~\ref{sec:partitioning}). Thus, it seems to possible to apply their technique, but replace their fixed point iteration \cite[Fig.~6]{DBLP:conf/pldi/GulwaniZ10} by one based on SMT-solving and path focusing, using acceleration if possible.

In recent years, because of improvement in SMT-solving,
techniques such as ours, distinguishing \emph{paths} inside loops,
have become tractable~\cite{Monniaux_POPL09,Beyer:2007:PI:1250734.1250769,Monniaux_LMCS10,Gawlitza_Monniaux_ESOP11}.
An alternative to using SMT-solving is to limit the number and length
of traces to consider, as in \emph{trace partitioning}
\cite{Rival_Mauborgne_TOPLAS07}, used in the Astr\'ee analyzer~\cite{Monniaux_ASIAN06,ASTREE_ESOP05,ASTREE_PLDI03}, but the criteria for limitation tend
to be ad hoc.
In addition, methods for abstracting the sets of paths inside a loop, weeding out infeasible paths, have been introduced~\cite{DBLP:conf/emsoft/BalakrishnanSIG09}.

With respect to optimality of the results, our method will generate the strongest inductive invariant inside the abstract domain if the domain satisfies the ascending chain condition and no widening is used; for other domains, like all methods using widenings, it may or may not generate it. In contrast, some recent works \cite{Gawlitza_Monniaux_ESOP11} guarantee to obtain the strongest invariant for the same analysis problem, at the expense of restriction to template linear domains and linear constructions inside the code.

\section{Conclusion and future work}
We have described a technique which leverages the bounded model checking capacities of current SMT solvers for guiding the iterations of an abstract interpreter. Instead of normal iterations, which ``push'' abstract values along control-flow edges, including control-flow splits and merges, we consider individual paths. This enables us, for instance, to use acceleration techniques that are not available when the program fragment being considered contains control-flow merges. This technique computes exact least invariants on some examples on which more conventional static analyzers incur gross imprecision or have to resort to syntactic heuristics in order to conserve precision.

We have focused on numerical abstractions. Yet, one would like to use similar techniques for heap abstractions, for instance. The challenge will then be to use a decidable logic and an abstract domain such that both the semantics of the program statements and the abstract values can be expressed in this logic. This is one direction to explore. With respect to the partitioning technique, \ref{sec:partitioning}, we currently express the partition as multiple explicit control nodes, but it seems desirable, for large partitions (e.g. according to Boolean values, as in B.~Jeannet's BDD-Apron library) to represent them succinctly; this seems to fit nicely with our succinct encoding of the transition relation as a SMT-formula.

Another direction is to evaluate the scalability of these methods on
larger programs. The implementation needs to be tested more to
evaluate the precision of our method on middle-sized
programs, the main advantage is that \aspic implements some of the
acceleration techniques. Analyzers such as \textsc{Astr\'ee} scale up
to programs running a control loop several hundreds of thousands of
lines long; translating such a loop to a SMT formula and solving for
this formula and additional constraints does not seem tractable. It is
possible that semantic slicing techniques
\cite{DBLP:conf/sas/Rival05} could help in reducing the size of the
generated SMT problems.


\newcommand{\doix}[1]{\endgroup}
\newcommand{\doi}{\begingroup\catcode`\_=13\def\_{\textunderscore}\doix}
\newcommand{\isbn}[1]{}
\newcommand{\issn}[1]{}
\newcommand{\biburlstyle}{\footnotesize~}

\bibliographystyle{plain}
\bibliography{sas_main}

\end{document}